\documentclass[twocolumn]{aastex61}
\pdfoutput=1 
\usepackage{amsmath,amstext}
\usepackage[T1]{fontenc}
\usepackage{times} 
\usepackage[figure,figure*]{hypcap}
\shorttitle{Tribocharging of KCl and ZnS}
\shortauthors{M\'endez Harper et al.}
\begin{document}
\title{Triboelectrification of KCl and ZnS particles in approximated exoplanet environments}
\author{Joshua M\'endez Harper}
\affiliation{School of Earth and Atmospheric Science, Georgia Institute of Technology, Atlanta GA, 30332, USA}
\affiliation{School of Electrical and Computer Engineering, Georgia Institute of Technology, Atlanta GA, 30332, USA}
\author{Christiane Helling}
\affiliation{School of Physics and Astronomy, University of St Andrews, St Andrews KY16 9SS, UK}
\affiliation{Centre for Exoplanet Science, University of St Andrews, St Andrews KY16 9SS, UK}
\author{Josef Dufek}
\affiliation{Department of Earth Sciences, University of Oregon, Eugene OR, 97403}
\begin{abstract}
When mobilized, granular materials become charged as grains undergo collisions and frictional interactions. On Earth, this process, known as triboelectrification, has been recognized in volcanic plumes and sandstorms. Yet, frictional charging almost certainly exists on other worlds, both in our own Solar System (such as Mars, the Moon, and Venus) and exosolar planets. Indeed, observations suggest that numerous planets in the galaxy are enshrouded by optically-thick clouds or hazes. Triboelectric charging within these clouds may contribute to global electric circuits of these worlds, providing mechanisms to generate lightning, drive chemical processes in the atmospheres, and, perhaps, influence habitability. In this work, we explore the frictional electrification of potassium chloride and zinc sulfide, two substances proposed to make up the clouds of giant exo-planets with >50x solar metallicities, including the widely-studied super-Earth GJ 1214b, super-earth HD 97658b,  Neptune-sized GJ 436b, and hot-Jupiter WASP-31b. We find that both materials become readily electrified when mobilized, attaining charge densities similar to those found on volcanic ash particles. Thus, if these worlds do indeed host collections of mineral particles in their atmospheres, these clouds are likely electrified and may be capable of producing lightning or corona discharge. 
\end{abstract}
\keywords{Triboelectrification --- extrasolar electricity --- salt clouds}
\section{Introduction}
Triboelectric charging, which broadly encompasses frictional and contact electrification, has been recognized in a wide variety of systems, both terrestrial and extraterrestrial, and likely operates beyond the confines of the Solar Systems. 
On Earth, perhaps the most dramatic, natural manifestation of the triboelectric effect are the displays of lightning observed during vigorous volcanic eruptions. These electrical storms have been recorded since antiquity and are the focus of important geophysical studies today \citep{pliny_the_younger_letters_1963, thomas_characterisation_2009, behnke_observations_2013, cimarelli_multiparametric_2016, mendez2018inferring}. Tribocharging has also been implicated in electrostatic processes within dust storms and dust devils \citep{stow1969dust, crozier1964electric, kamra1972measurements, bo2013field}. On smaller scales, bees and other insects may rely on the electrostatic forces generated during fricional electrification to successfully pollinate plants--including the crops billions of humans require to survive \citep{vaknin2000role, corbet2014buzz}. Within industrial and other man-made environments, granular materials flowing through pipes and hoppers often charge frictionally, generating significant potentials which represent shock or explosion hazards \citep{matsusaka2000electrification, hendrickson_electrostatics_2006, lacks_contact_2011}. While direct measurements are lacking, analogous granular charging processes are believed to operate in a number of extraterrestrials settings. On Mars, for instance,  dust devils likely contain abundant charged grains and may host small-scale spark discharges or corona \citep{krauss_experimental_2003, farrell_is_2015, mendez2017imulating}. Additionally, recent experiments have shown that stormy winds on Titan could triboelectrify organic particles sufficiently to cause aggregation, increasing saltation thresholds across the Saturnian moon's extensive dune fields \citep{mendez_harper_electrification_2017}.
 
It is important to mention that while triboelectric charging is "the oldest manifestation of electricity known to man, it still remains today quite obscure as to the mechanisms active" \citep{kunkel1950static}. Indeed, the nature of the charge carriers being exchanged when two surfaces come into contact or when these two are rubbed together has not been resolved \citep{lacks_contact_2011}. One model--the trapped electron model--suggests that particles become charged as electrons caught in unfavorable energy states on one surface migrate to low energy states on another particle during particle-particle collisions \citep{lowell1980contact, lowell_triboelectrification_1986, lowell_triboelectrification_1986-1}. Such model has gained considerable acceptance because it not only accounts for electrification in systems of chemically identical grains, but can explain the oft-reported tendency of smaller grains to charge negatively while larger ones gain positive charge (i.e. the size-dependent bi-polar charging of chemically-identical substances) \citep{forward_charge_2009, lacks_contact_2011, bilici_particle_2014}. More recently, however, experimental and numerical investigations have suggested that the same charging behavior can be achieved through the partitioning and concentration of water ions (H\textsuperscript{+} and OH\textsuperscript{-}) on particles of differing sizes \citep{gu_role_2013, xie_effect_2016}.

\begin{figure}
        \includegraphics[width=\linewidth, clip]{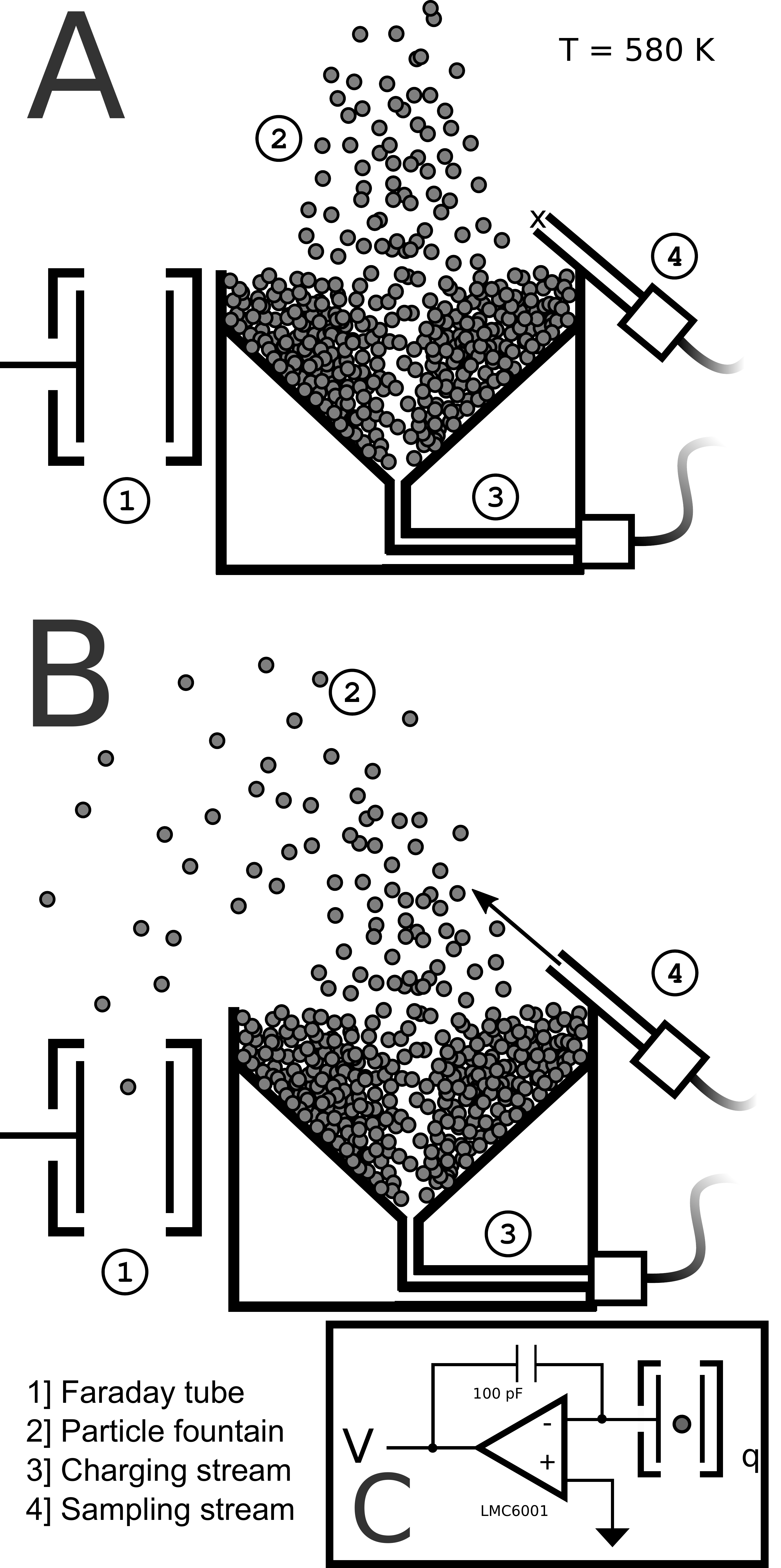}
        \caption{Experimental apparatus. Particles are charged in fluidized bed that generates many particle-particle collisions, while minimizing interactions with foreign surfaces. The design has been described in detail in \protect \cite{forward_triboelectric_2009}, \cite{bilici_particle_2014} and \cite{mendez_harper_effects_2016}. Panel A shows the operation during the charging period. Here, the particle bed is electrified by mobilizing grains with a narrow stream of dry air injected into the base of the bed. This charging process lasts around 20 minutes. Panel B shows the system during the measurement period. While still being mobilized, a second stream of air blows particles out of the granular fountain. Ejected grains fall into a micro-machined through-type Faraday cup, where their charge is measured. The Faraday cup sensor can measure charges down to 1 femtocoulomb (or around 6500 electrons).}
        
        \label{fig_10_1}
\end{figure}

Despite these outstanding (certainly, fundamental) questions, triboelectric charging appears to be an inherent characteristic of mobilized granular flows and, thus, is likely also present within large granular reservoirs on exosolar objects. Recent, observations of distant solar systems have revealed a surfeit of worlds with featureless transmission spectra  \citep{berta2012flat, kreidberg2014clouds, knutson2014featureless, morley2015thermal, sing2016continuum, gibson2017vlt} or other signatures that have been interpreted to be the effects of high-elevation clouds and hazes. These planets include super-Earths GJ 1214b \citep{kreidberg2014clouds} and  HD 97658b \citep{knutson2014hubble}, Neptune-sized GJ 436b \citep{knutson2014featureless} and GJ 3470b \citep{ehrenreich2014near, dragomir2015rayleigh}, and hot-Jupiters HD 189733b \citep{pont2013prevalence} and WASP-31b \citep{gibson2017vlt}.  The extreme conditions on some of these worlds suggest that their clouds are composed not of water droplets, but of mineral dust particles or organic hazes \citep{pont2008detection, sing2009transit, moses2013compositional, kreidberg2014clouds, charnay20153d, mbarek2016clouds, lee2018dust}. On GJ 1214b, for instance, clouds may be composed of condensed (solid) salts, such as potassium chloride (KCl) or zinc sulfide (ZnS). Such phases may also be present on super-earth HD 97658b \citep{mbarek2016clouds},  Neptune-sized GJ 436b \citep{moses2013compositional}, and hot-Jupiter WASP-31b (e.g. \cite{sing2014hst}). Organic hazes, similar to those found in the atmosphere of Titan, have also been invoked to account for flat transmission spectra \citep{kreidberg2014clouds, horst2018haze}. 

Regardless of their exact composition, particles suspended in these exoplanet environments likely undergo repeated particle-particle collisions in response to atmospheric circulation. Such dynamics have been inferred to drive efficient triboelectrification (e.g. \cite{helling_ionization_2013}), resulting in electrified cloudy or hazy environments at elevation. As happens within Earth's clouds, exoplanet clouds are likely gravitationally stratified, meaning that smaller, lighter grains become concentrated at the top of the clouds, while larger, heavier grains remain at lower elevation (see figures 2 and 8 in \cite{helling2008dust}; also \cite{helling2013modelling, helling2016ionisation}). Because, as discussed above, the polarity of charge collected by particles from triboelectric processes depends on grain size, this stratification (smaller particles at elevation; larger particles on the bottom) has the ability to set up coherent electric fields. Such charge separation occurs in both thunderstorms and volcanic plumes and ultimately drives spark discharges--lightning--either through conventional breakdown of the gas or via runaway electron avalanche \citep{kikuchi_atmospheric_1982, gurevich_runaway_1992, james1998specialvolcanic, dwyer_initiation_2005, dwyer_physics_2014, cimarelli_experimental_2014, aizawa_physical_2016}. On Earth, these discharges support a global electric circuit and have the ability to modulate chemical and physical reactions in the atmosphere \citep{price1993global, rakov2007lightning, siingh2007atmospheric, genareau2015lightning, wadsworth2017size, mueller2018first}. Indeed, lightning may have been involved in the production of prebiotic molecules in an early-Earth environment \citep{miller1959organic, navarro1998nitrogen} and such lightning has been hypothesized to have been associated within dusty flows (namely, volcanic plumes) rather than hydrometeor clouds \citep{navarro1998nitrogen, segura2005nitrogen, johnson2008miller}. If the mineral clouds inferred to exist on extrasolar worlds can be considered analogs to the dusty environments in our own solar systems, charging within these systems may also stimulate a wide array of electrostatic phenomena and  help catalyze prebiotic chemistry \citep{hodosan2016lightning}.  

In the present work, we characterize the triboelectric behavior of two substances inferred to compose some exoplanet clouds: KCl and ZnS. We conduct our experiments under approximated GJ 1214b conditions, but our results can be analyzed in the context of other exoplanets as well. We show that both materials become charged when fluidized and attain charge densities comparable to those observed on ash particles falling out of volcanic plumes and thunderstorm water droplets. Furthermore, based on the charge densities measured on particles, we determine the conditions (particle size, volume fraction, and atmospheric pressure) required to generate discharges in a hypothetical cloud. Additionally, we compare our results with previous numerical models utilized to explore charging in mineral clouds (e.g. \cite{helling_ionization_2013}).

\section{Methods}
Triboelectrification of KCl and ZnS grains was achieved using an apparatus similar to that previously described in \cite{forward_particle-size_2009} and \cite{mendez_harper_effects_2016}. The device consists of a spouted bed contained within an environmental chamber (figure \ref{fig_10_1}a show the apparatus in schematic form). The system involves a machined steel cup (width = 6 cm, depth = 6 cm) capable of holding ~100 g of particles. This particle bed is excited by injecting a stream of air through a 500-micron hole in the base of the cup. The air stream fluidizes the particle bed (meaning, particles are mobilized into a fountian-like structure; see figure \ref{fig_10_1}), causing grains to undergo frictional interactions. The geometry of the cup was selected to generate copious particle-particle collisions, while minimizing particle-wall contacts. Thus, triboelectric charging in this work manifests overwhelmingly from the interaction of chemically-identical surfaces. 
\begin{figure}
        \includegraphics[width=\linewidth, clip]{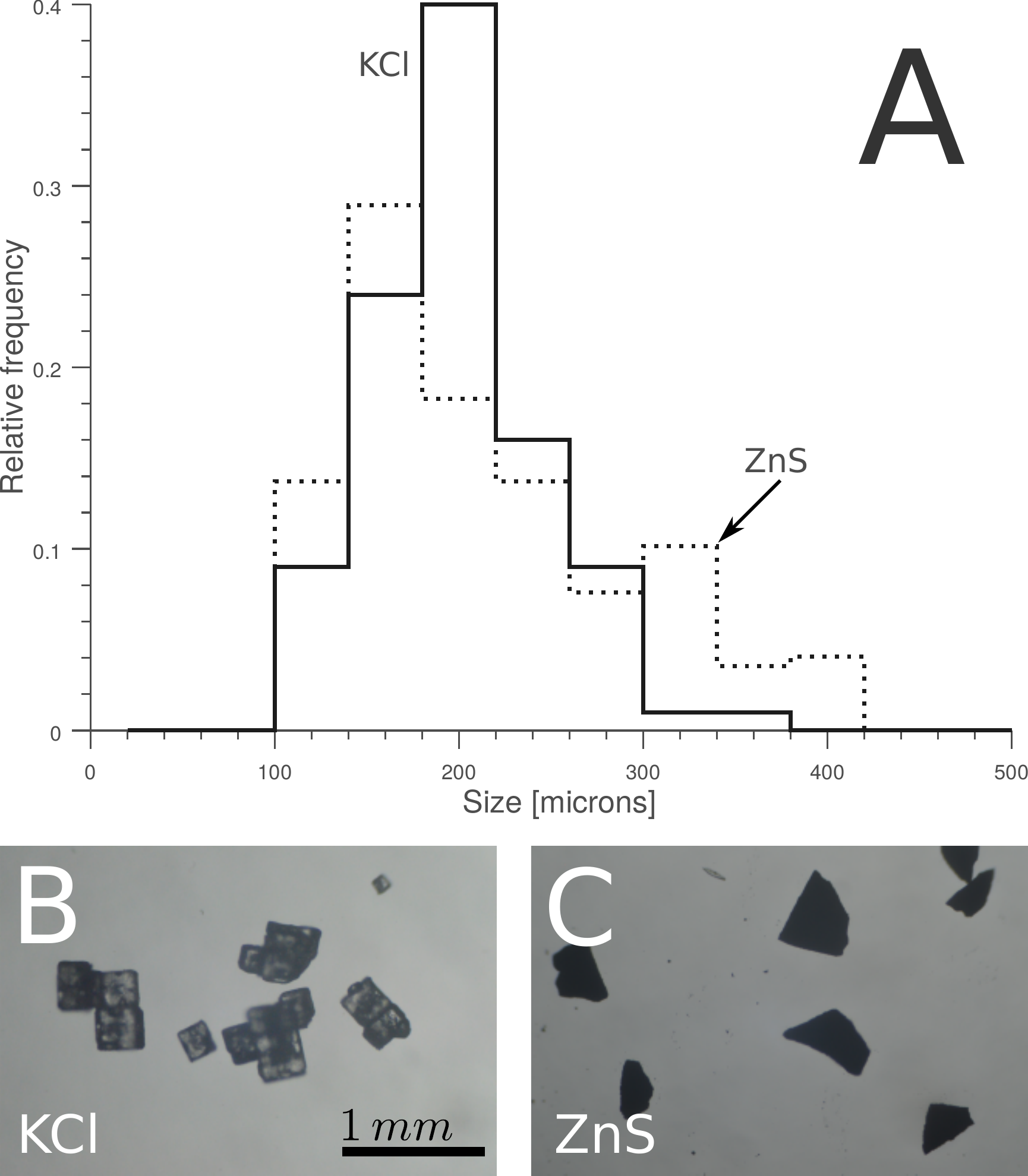}
        \caption{A) Grain size distributions (spherical-particle equivalents) used in these experiments for KCl (solid curve) and ZnS (dashed curve). B) Micrograph of KCl particles. C) Micrograph of ZnS particles.}
        \label{fig_10_2}
\end{figure}    
        
\begin{figure}
        \includegraphics[width=\linewidth, clip]{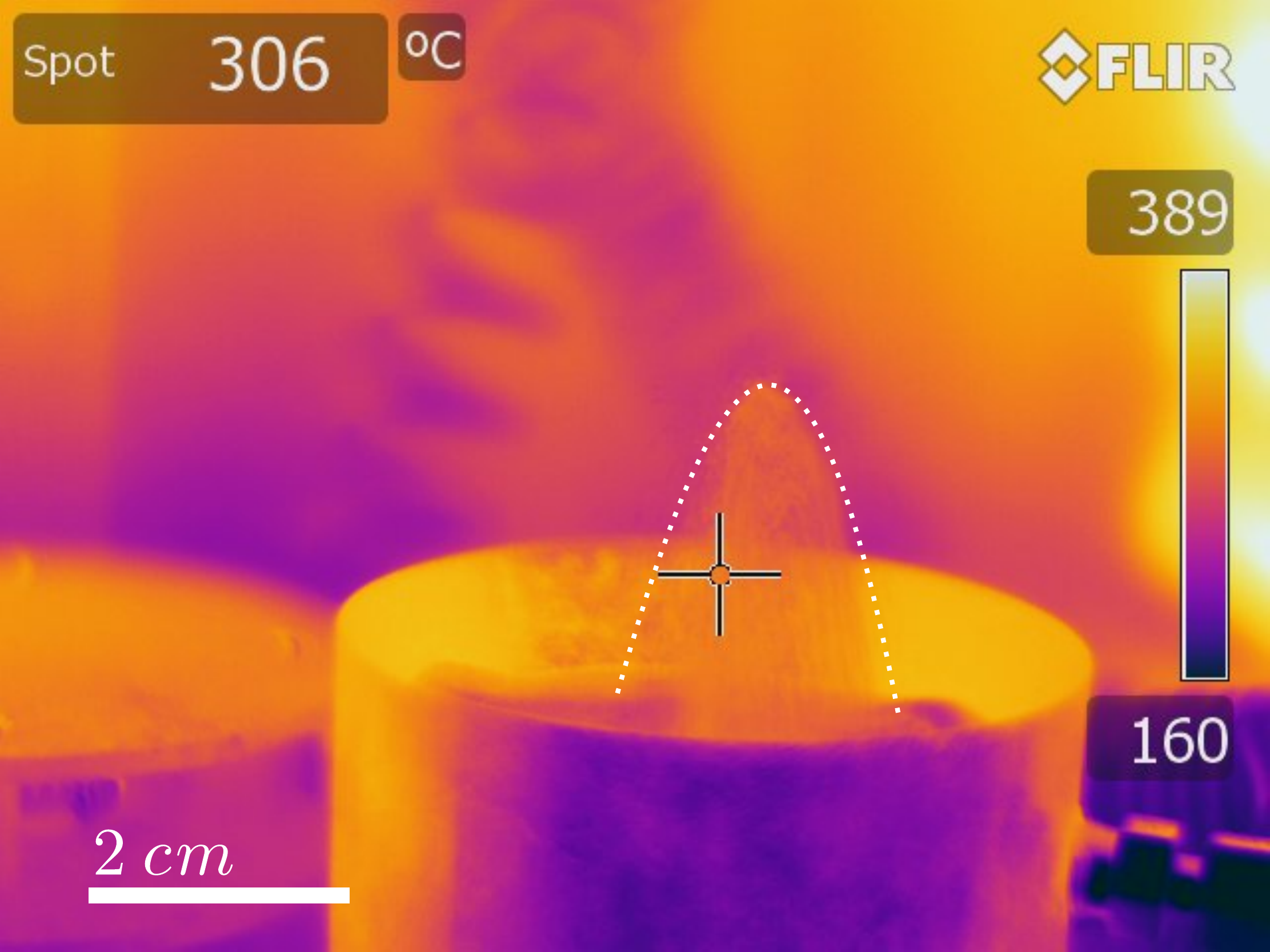}
        \caption{Thermal infrared image of the experimental apparatus in the furnace. Spot temperature of the fountain (outlined by dotted curve) is ~300 \textsuperscript{o}C (580 K)}
        \label{fig_10_3}
\end{figure}

We used commercially available KCl (Sigma-Aldrich, 60130) and ZnS (Sigma-Aldrich, 14459) powders in our experiments. The materials were sieved to obtain nominal size distributions in the range of 125-500 $\mu$m (nominal). True size distributions are rendered in figure \ref{fig_10_2}. The fluidizing apparatus was placed in a closed-loop controlled furnace where the temperature was maintained at 550-600 K . This temperature range is congruent with the cloud-forming temperatures for planets like GJ 1214b, HD 97658b, GJ 436b, and GJ 3470 b \citep{lewis2010atmospheric, demory2013spitzer, van2014transit} and will not lead to phase transition of the materials considered. Additionally, the pressure within furnace was maintained at 1 bar (101 kPa) and relative humidity was kept $<$1\%. Once the experimental setup equilibrated with the surrounding atmosphere (~4 hours), a valve was opened allowing a jet of air to pass through the hole in the base of the cup, resulting in a particle fountain ~3-4 cm high (Figure \ref{fig_10_3}). Fountaining under these conditions was allowed to proceed for 20 minutes, a period long enough to ensure that particles reached an electrostatic steady state \citep{forward_charge_2009, mendez_harper_effects_2016}. At the conclusion of the charging period, a second air stream was activated at 45 degrees to the base of the fountain, catapulting suspended grains out of the fountain and into a micro-machined through-type Faraday cup (TTFC; see Figure \ref{fig_10_1}b and \ref{fig_10_3}). The aperture of the Faraday cup is 1 mm in diameter, allowing only a few particles to traverse into its interior at a time. Upon entering the cup, a particle produces a current that flows from the cup's interior electrode to a charge amplifier built around an LMC6001 operation amplifier. The amplifier then generates a voltage pulse proportional to the total charge on the particle. In this manner, we were able to characterize the charge on individual particles with a resolution of 1 femtocoulomb (fC; on the order of 1e\textsuperscript{3} electrons). For each substance, we measured the charge on 700-800 particles.

While the exoplanets discussed above have atmospheres which are either of solar composition (H\textsubscript{2} or He) or perhaps water-rich, the use of dry air in our experiments was justified by the analysis presented in \cite{helling_ionization_2013}. These investigators suggest that electrostatic processes on exoplanets are not largely dependent on gas composition. While we continue their reasoning in our experiments, we do so with caution given that the effect of environmental conditions on triboelectric charging are poorly constrained. Indeed, some experiments have found no correlation between granular electrification and gas composition \citep{merrison2012factors} while others have reported clear dependencies \citep{matsuyama1995charge,mendez_harper_effects_2016}. Until such questions are resolved definitely, the results presented here should be considered preliminary and of first order.
        
\begin{figure}
        \includegraphics[width=\linewidth, clip]{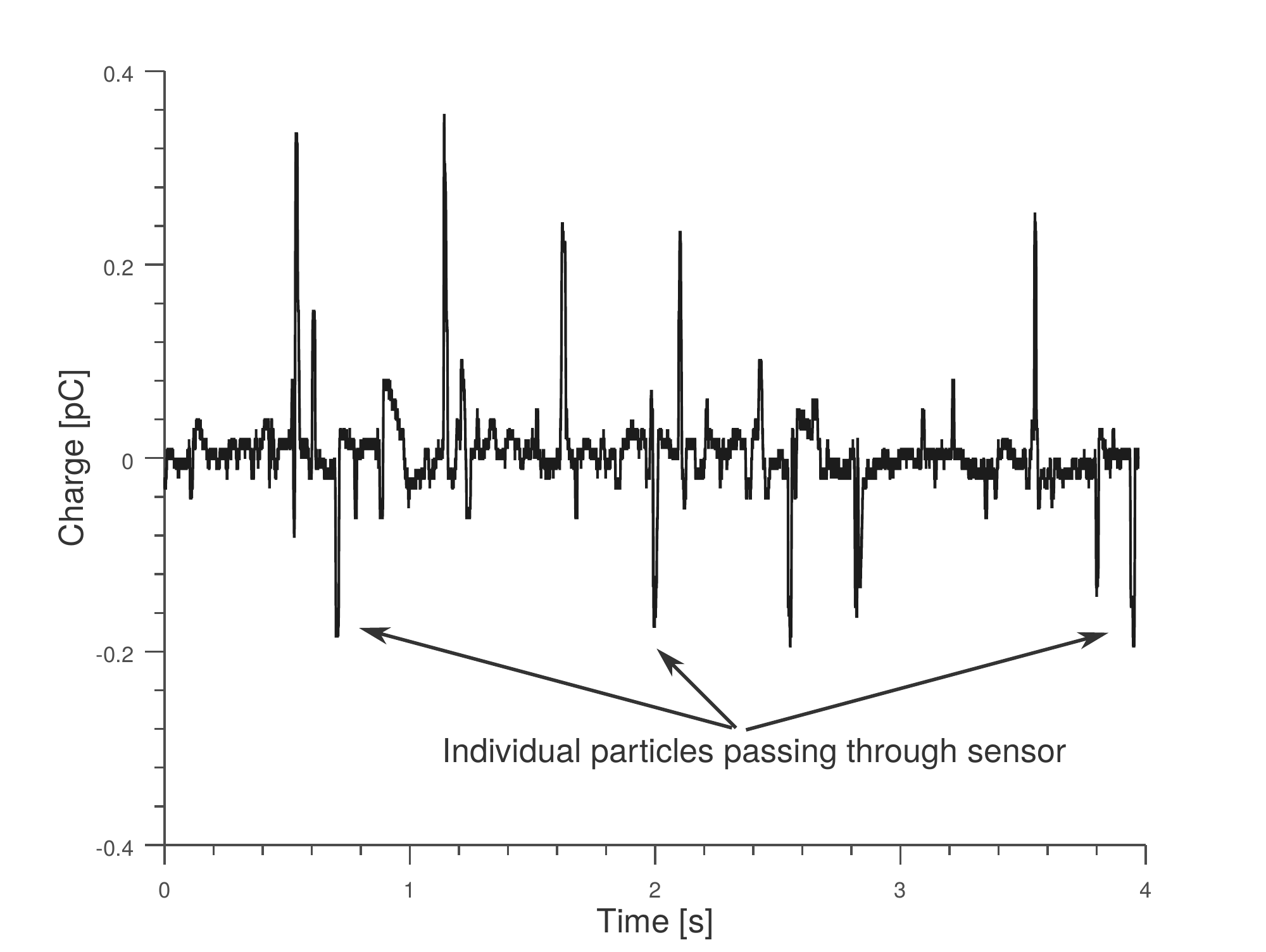}
        \caption{Raw charge data collected from the output of the charge amplifier. Each pulse represents the passage of a single particle through the Faraday tube.}
        \label{fig_10_4}
\end{figure}    
\section{Results}

\begin{figure}
        \includegraphics[width=\linewidth, clip]{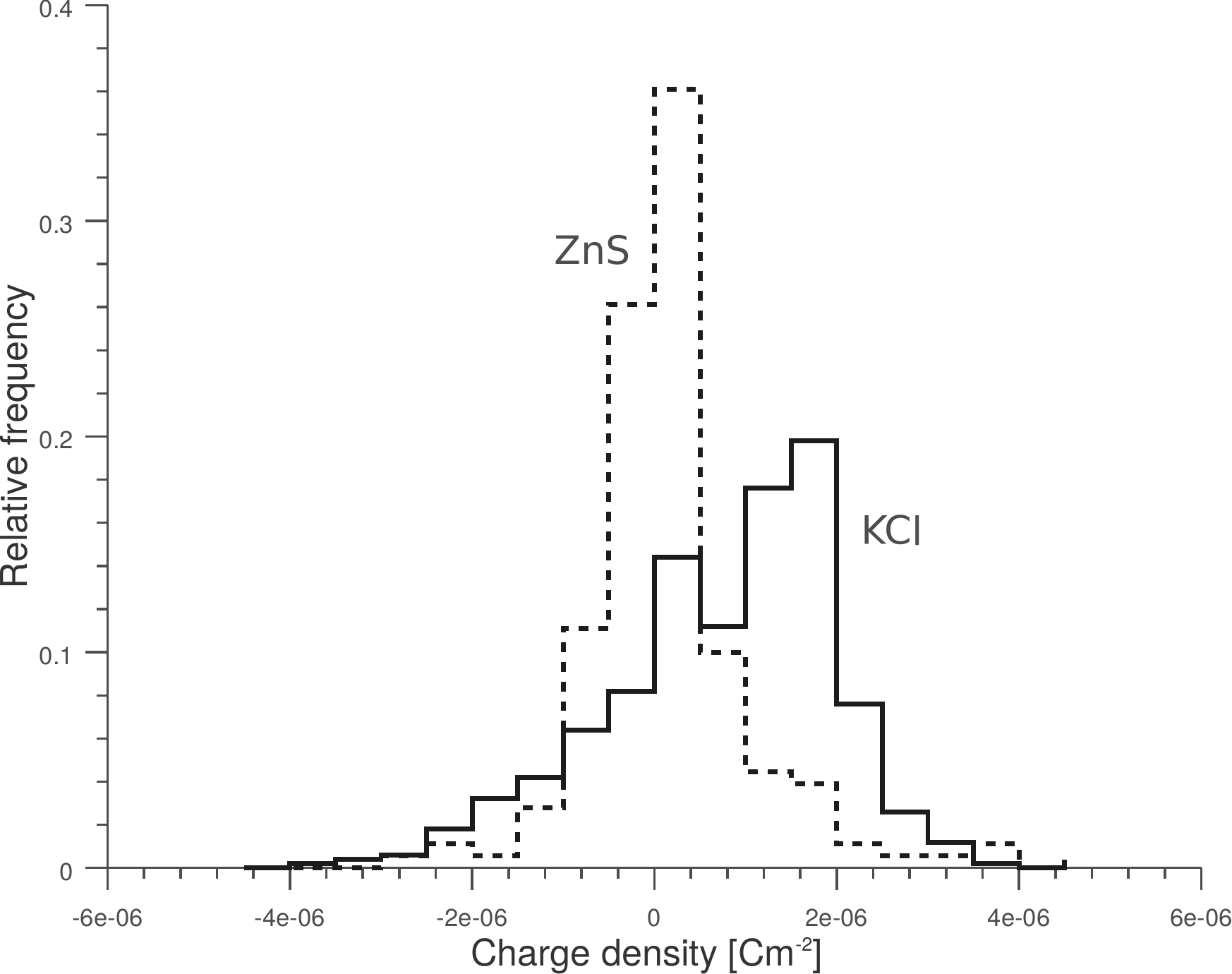}
        \caption{A) Mean charge density distributions for KCl (solid) and ZnS (dashed). Note that while both materials have comparable maximum charge densities, KCl has an overall broader distribution that ZnS. The higher chargeability of observed in the KCl flow may be attributed to particle shape. While KCl grains are cubic, ZnS particles are much more angular. Charge on ZnS may be lost at sharp particle tips, reducing the charge densities on grains (\ref{fig_10_2}b and c).}
        \label{fig_10_5}
\end{figure}

Raw data from an experiment with KCl particles are shown in Figure \ref{fig_10_4}. Each pulse represents a single particle traversing the TTFC. Particles of both materials attained maximum charges of several 100 to 1000 fC during the fluidization process. Note that particles charged both negatively and positively, satisfying charge conservation. A standard way of assessing the chargeability of a material is to analyze its \textit{charge density} or its charge normalized by its surface area. Here, charge densities for both KCl and ZnS samples were obtained by dividing the particle  charge distributions by the average spherical-equivalent particle area (although, particles are significantly \textit{non}-spherical, spherical equivalents are commonly used in this form of calculations). The average charge densities for both KCl and ZnS range between 10\textsuperscript{-9} - 10\textsuperscript{-6} Cm\textsuperscript{-2} (See Figure \ref{fig_10_5}a) and are comparable to the ranges of charge densities observed on volcanic ash in laboratory experiments (10\textsuperscript{-9} - 10\textsuperscript{-5} Cm\textsuperscript{-2}; \cite{james_volcanic_2000, mendez_harper_effects_2016, mendez_harper_electrification_2017}) and those measured on submillimeter-sized pyroclasts falling out of volcanic plumes (i.e. \textit{in-situ} and field measurements; 10\textsuperscript{-6} - 10\textsuperscript{-5} Cm\textsuperscript{-2}; \cite{gilbert_charge_1991, miura2002measurements}). While both substances attain similar maximum charge densities, the charge density distribution for KCl is broader than that of ZnS, meaning that a greater number of KCl particles held elevated charge densities. This discrepancy may be attributed to differences in particle morphology. While KCl grains are cubic in shape, ZnS grains are much more angular (Figure \ref{fig_10_2}b). Sharp tips on these grains may promote charge loss to the surrounding gass resulting in overall smaller charge densities on ZnS grains. 

\begin{figure}
        \includegraphics[width=\linewidth, clip]{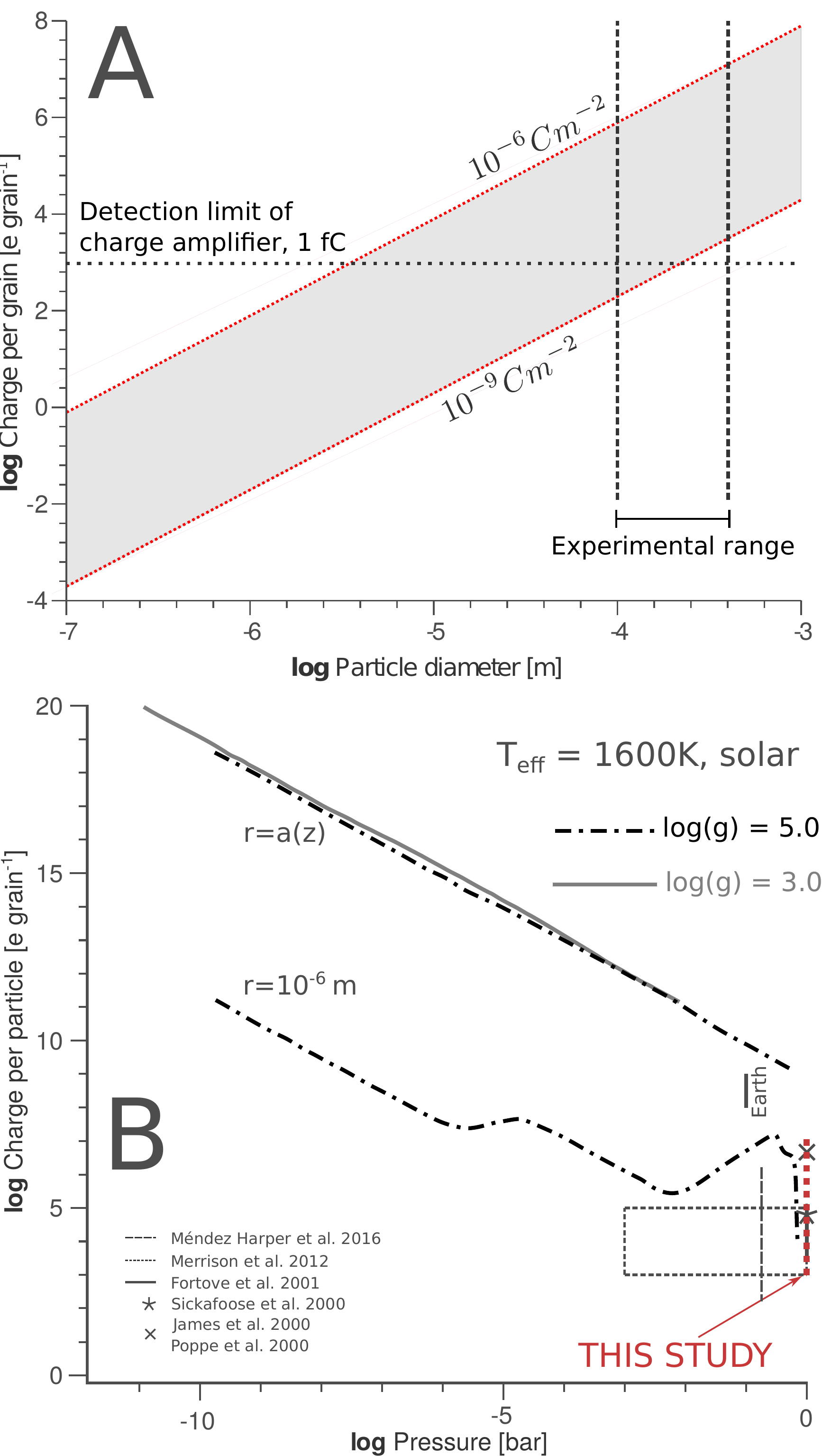}
        \caption{A) From the charge density distributions, the absolute charge per particle can be extrapolated for a range of particle sizes. In the context of our experiments, we expect each particle to carry between 100 and 1e7 elemental charges.  B) The charges observed in our experiments are comparable to those measured by other investigators on volcanic ash, sand, and other particles. Plotted on the same graph are the estimates for charge per particle required to produce breakdown with an exoplanet mineral cloud. Figure modified from \cite{helling_ionization_2013}.}
        \label{fig_10_6}
\end{figure} 

From the surface charge density one can compute the charge per grain as a function of particle size. This relationship is shown in figure \ref{fig_10_6}a for particles with diameters in the range of 100 nanometers to 100 microns. To put these numbers into context with other studies, we overlay our data onto figure 7 from \cite{helling_ionization_2013} which  displays the number of charges per grain required for discharges in a hypothetical mineral cloud along with charge densities measured in several experimental works (figure \ref{fig_10_6}b). Our data is generaly congruent with previous measurements involving silicate materials (sand and volcanic ash also at 1 bar) and supports the numerical analysis presented in \cite{helling_ionization_2013}.
   
\section{Discussion}
In dusty atmospheres, breakdown processes exist on a wide range of scales and are controlled by both microphysical and macroscopic mechanisms \citep{behnke_changes_2015, farrell_is_2015, cimarelli_multiparametric_2016, mendez2018inferring}. Spark discharges between two charged surfaces (two capacitor plates, for example) may occur when the electric field between these surfaces exceeds the dielectric breakdown of air, creating a conducting channel that carries charge between the two surfaces. Lightning is a natural example of a spark discharge \citep{rakov2007lightning}. The presence of a single, highly charged object in a fluid may also produce a discharge by creating a region of plasma in the vicinity of its surface. In this case, charge is lost directly to the surrounding gas \citep{cross1987electrostatics}. \textit{Partial discharges} of this sort often create visble, crown-like glows around the charged object and are known as \textit{corona}. In his magnum opus \textit{Moby Dick}, American novelist Herman Melville describes corona discharge, known to sailors as corpusants or Saint Elmo's light, as a "pallid fire" buring from the whaler's tall masts during a night of foul weather \citep{melville1994moby}. In more modern contexts, partial discharge may complicate the operation of some systems--such high-voltage power lines where ionization generates power loses \citep{bartnikas_detection_1990}--, but many benefits may also be drawn from corona processes like the removal of unwanted static charge on industrial equipment \citep{jay1972process}. 

In \cite{helling_ionization_2013}, the authors place constraints on the charge required on particles to produce discharges given certain extrasolar environments. Here, we ask the converse question: given the charge densities measured on KCl or ZnS grains experimentally, what environmental conditions lead to breakdown? The plates of an imaginary capacitor yield a zeroth order solution. The maximum voltage, $V_b$, that can be sustained between the two charged surfaces separated by some distance, $d$, is dependent on the composition of the enveloping gas and its pressure \citep{paschen_ueber_1889}. This relation is known as Paschen's law:
\begin{equation} \label{eq10.1}
V_b =  \frac{Bpd}{\text{ln}({\frac{Apd}{\text{ln}(1/\gamma + 1)}})} 
\end{equation}

In equation \ref{eq10.1}, $p$ is pressure, $A$ and $B$ are constants which give Townsend's ionization coefficient, and $\gamma$ is Townsend's second ionization coefficient with a value of 0.01 \citep{kok_wind_blown_2009}. These constants are listed for a number of gases in Table \ref{tab_10_1} \citep{raizer_gas_2001, helling_ionization_2013}. For the present application, however, it is more useful to describe the dielectric strength of the atmosphere in terms of a breakdown electric field $E_b$ rather than a voltage \citep{kok_wind_blown_2009}:
  
\begin{equation} \label{eq10.2}
E_b =  \frac{Bp}{\text{ln}({\frac{Apd}{\text{ln}(1/\gamma + 1)}})} 
\end{equation}

\begin{figure}
    \centering
        \includegraphics[width=\linewidth, clip]{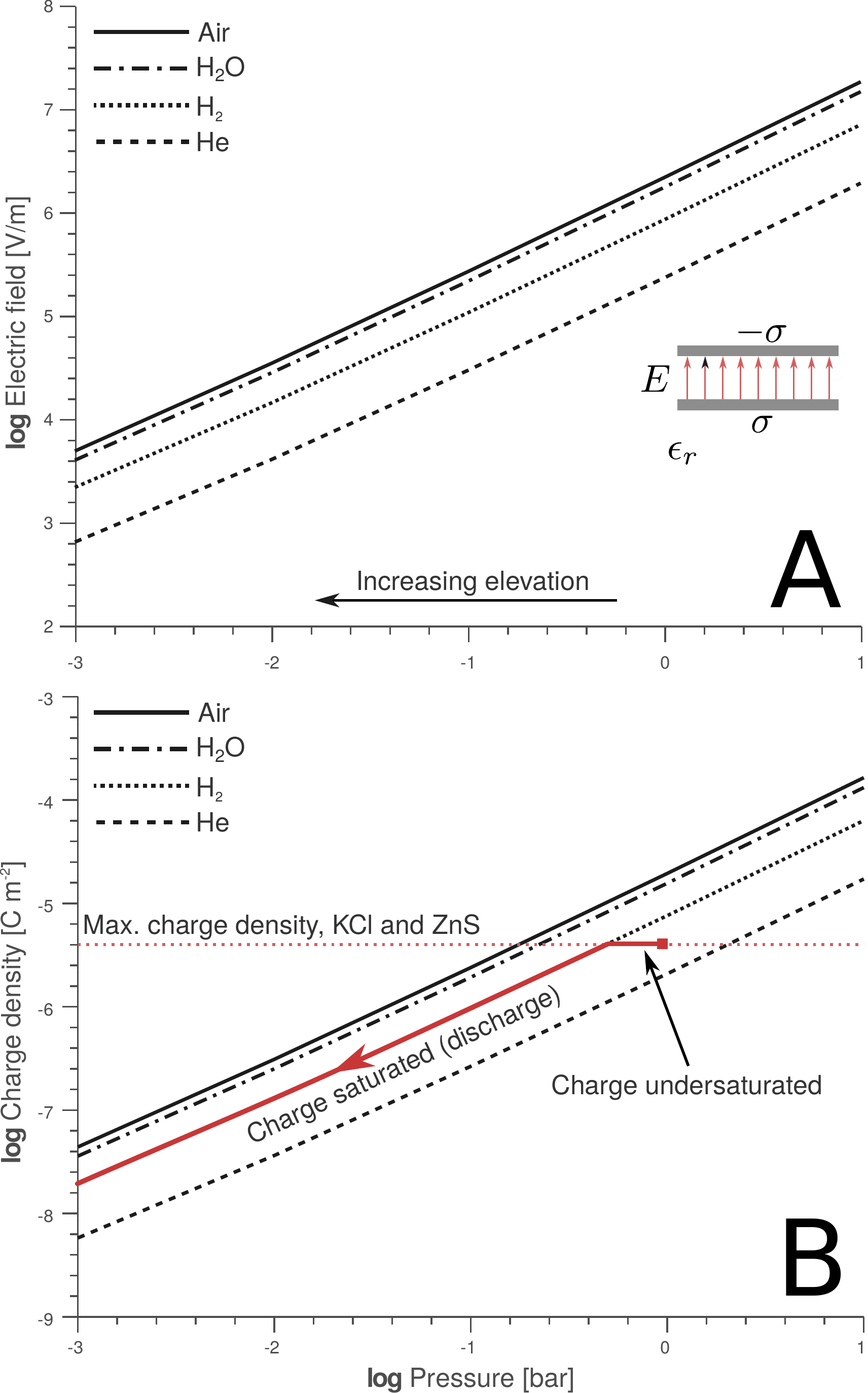}
        \caption{A) Maximum electric field across a 1 m spacing for air (solid), H\textsubscript{2}O (dashed-dotted), H (dotted), and H\textsubscript{2} (dashed) as a function of pressure. B) Maximum charge density for a surface in air (solid), H\textsubscript{2}O (dashed-dotted), H (dotted), and H\textsubscript{2} (dashed) as a function of pressure. The shaded area, topped off by dotted red curve, is representative of the range of charge densities measured in our experiments. The thick, solid red curve staring with a square represents the electrostatic narrative of a charged particle as it is advected to elevations with lower pressure in an H\textsubscript{2} atmosphere. To satisfy Paschen's law, the grain's surface charge density must decrease with altitude.}
        \label{fig_10_7}
\end{figure}

From equation \ref{eq10.2}, it follows that the breakdown field on Earth (at a pressure at 1 bar or 101 kPa) across a distance of 1 m is on the order of 10\textsuperscript{6} Vm\textsuperscript{-1}. Exceeding this field causes the gas to conduct. For H\textsubscript{2} and He, E\textsubscript{b} = 8 $\times$ 10\textsuperscript{5} Vm\textsuperscript{-1}  and E\textsubscript{b} = 2.5 $\times$ 10\textsuperscript{5} Vm\textsuperscript{-1}, respectively, while a water-rich atmosphere would breakdown close to 1.8 $\times$ 10\textsuperscript{6} Vm\textsuperscript{-1}. For each of these three gases (H\textsubscript{2}, He, H\textsubscript{2}O) and air, the variation in breakdown field with pressure is rendered in figure \ref{fig_10_7}a (for $d$ = 1 m). Note that for all pressures, helium is the weakest gas (i.e. it breaks down at lower electric fields for a given pressure), while air is the strongest. Across these atmospheric compositions, $E_b$ has a variation close to an order of magnitude.

\begin{table}
\caption{Paschen curve parameters for a number of gases \citep{raizer_gas_2001, helling_ionization_2013}.}
\begin{center}
\begin{tabular}{ccc}
\textbf{Gas} & \textbf{A (m\textsuperscript{-1} Pa\textsuperscript{-1})} & \textbf{B (V m\textsuperscript{-1} Pa\textsuperscript{-1})} \\
\hline
H\textsubscript{2} & 3.75 & 97.50   \\
He & 2.25 & 25.50  \\
H\textsubscript{2}O & 9.75 & 217.50  \\
Air & 11.25 & 273.77  \\
\label{tab_10_1}
\end{tabular}
\end{center}
\end{table} 

Knowling the breakdown field, one can compute the maximum charge density, $\sigma_b$, that can be supported on the two surfaces of the hypothetical capacitor \citep{hamamoto_experimental_1992}:

\begin{equation} \label{eq10.3}
\sigma_{b} = \epsilon_r \epsilon_o E_{b}
\end{equation}

where $\epsilon_r$ is the relative dielectric constant of the gas and $\epsilon_o$ is the permittivity of free space. For Earth (at 1 bar, 101 kPa), this expression yields 2.7 $\times$ 10\textsuperscript{-5} Cm\textsuperscript{-1}. In hydrogen, helium, and water atmospheres (again, at 1 bar), the breakdown charge densities are 7.6 $\times$ 10\textsuperscript{-6} Cm\textsuperscript{-1}, 2.0 $\times$ 10\textsuperscript{-6} Cm\textsuperscript{-1}, 1.6 $\times$ 10\textsuperscript{-5} Cm\textsuperscript{-1}, respectively. The variation of the maximum charge density with pressure is rendered in figure \ref{fig_10_6}b for the four gases. On this same plot, we superimpose the range of charge densities observed in our experiments which have maximum magnitudes of ~4 $\times$ 10\textsuperscript{-6} Cm\textsuperscript{-2} (for both KCl and ZnS; grey shaded area in figure \ref{fig_10_7}b. Note that for all gases except He, charge densities of 10\textsuperscript{-6} Cm\textsuperscript{-2} on hypothetical capacitor plates separated by a distance of 1 m would be insufficient to initiate discharge processes at 1 bar. However, these charge densities may lead to discharge if the pressure of the gas surrounding the host surfaces decreases. In the particular case of GJ 1214b, theoretical considerations and numerical models have suggested that salt clouds are generated within the atmosphere at pressures close to 1-0.1 bar, but are then lofted to elevations where the pressures range between 10\textsuperscript{-2} and 10\textsuperscript{-3} bar \citep{kreidberg2014clouds, charnay20153d}. Thus, surfaces that achieve charge densities comparable to those in our experiments while at depth likely become charge-saturated when they are advected to more rarefied regions. To remain at or below the Pascehn limit, up-lifted surfaces shed charge through either spark or partial discharge and their charge density decreases with altitude. This processes is rendered in figure \ref{fig_10_7}b for an H\textsubscript{2} atmosphere (red curve). Other, work also indicates that particles may also rain out of clouds, providing a mechanism for further electrification \citep{helling2008dust,  helling2013modelling}.

\begin{figure}
    \centering
        \includegraphics[width=\linewidth, clip]{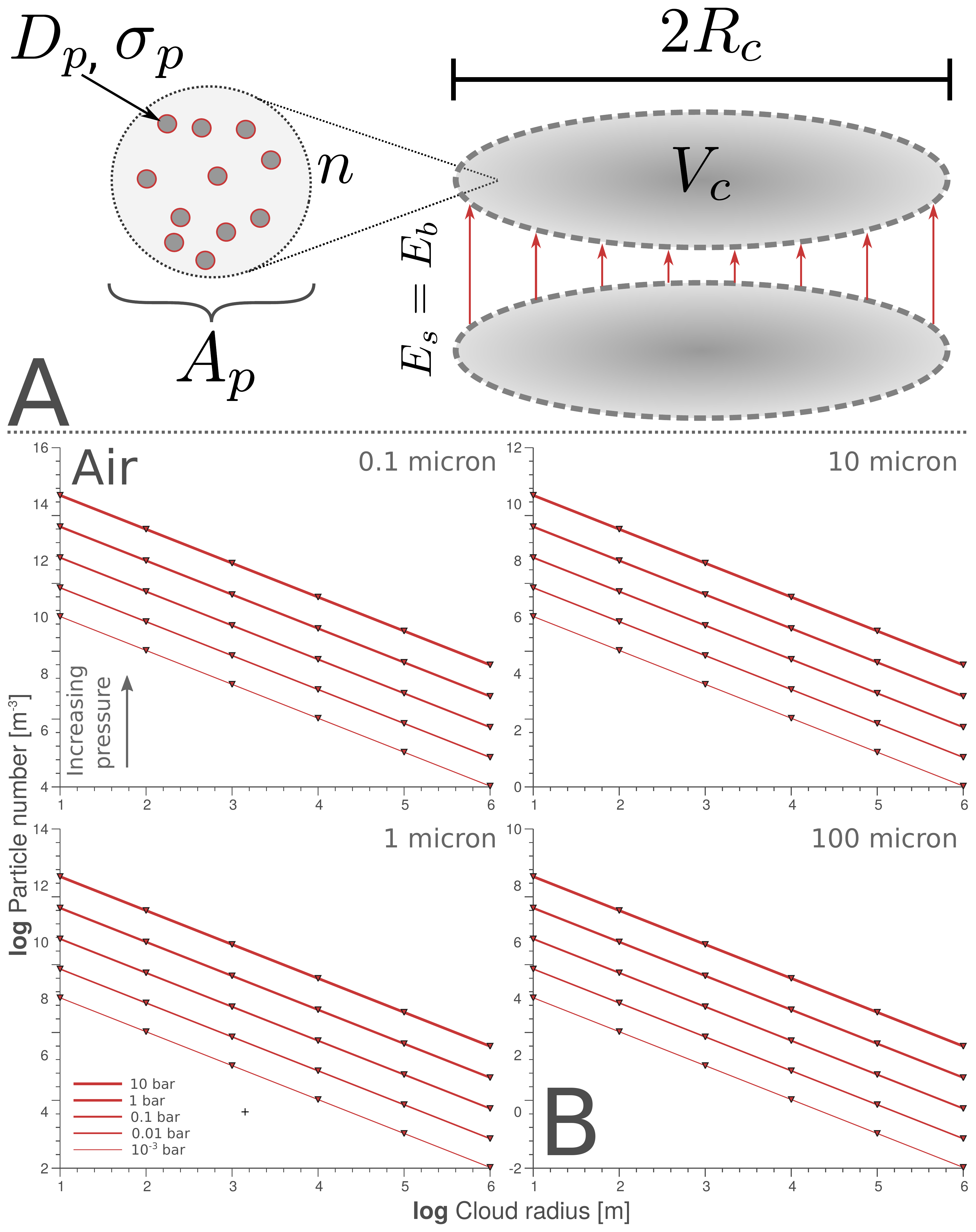}
        \caption{A) Schematic of hypothetical mineral cloud arrangement used to determine breakdown conditions for a number of atmospheric conditions and the surface charge densities measured experimentally. B) Particle number density required to produce the large-gap electric field at the cloud edge as a function of particle size, atmospheric pressure, and cloud radius. The calculation assume a uniform charge densities on particles on the order of 10\textsuperscript{-6} C m\textsuperscript{-2} (the maximum charge densites observed in our experiments).}
        \label{fig_10_8}
\end{figure}

In a cloud, however, charge is not distributed across a continuous surface (as would be the case for the capacitor plate we have been using as an analogy), but among many discrete particle surfaces. Rather than a pair of plates, consider two spherical clouds, each with radius $R_c$, surface area $S_c$, and volume $V_c$ (see figure \ref{fig_10_8}a). To generate discharge conditions at a cloud's surface, the superposition of the electric fields associated with individual grains must produce a total effective electric field, $E_s$, that exceeds the breakdown value $E_b$ (note that $E_b$ is the wide-gap breakdown field--i.e. when $d \approx$ 1 m):
\begin{equation} \label{eq10.4}
\oint E_sdS_c = \frac{Q(n, A_p, \sigma_p)}{\epsilon_r \epsilon_o}.
\end{equation}
Above, $Q$ is the total charge in the cloud which depends on the number of grains per unit volume $n$, the surface area per particle, $A_p$, and the particle surface charge density $\sigma_p$:
 
\begin{equation} \label{eq10.5}
Q = \sigma_p A_p V_c n.
\end{equation}
Solving equation \ref{eq10.4} (which is nothing more than a rendition of Gauss's Law) for the particle charge density and setting $E_S = E_b$, yields:
\begin{equation} \label{eq10.6}
\sigma_p = \frac{3E_b (p) \epsilon_r \epsilon_o}{\pi D_p^2 n R_c}.
\end{equation}
Equation \ref{eq10.6} shows that the maximum charge density that can be supported on an individual particle in the cloud $\sigma_p$ may be smaller or larger than that given by equation \ref{eq10.3} ($\sigma_b$) depending on the size and abundance of grains. Indeed, $\sigma_p$ will be smaller than $\sigma_b$ if $3/{\pi D_p^2 n R_c} < 1$. In other words, the discharge threshold becomes easier to attain in spatially extensive clouds with a high number density of large particles. Such conclusion makes intuitive sense; to generate a given electric field at the cloud surface, a cloud with many, large particles within a given volume requires constituent particles to carry less charge per grain than a small, dilute cloud with few small particles.

\begin{figure}
    \centering
        \includegraphics[width=\linewidth, clip]{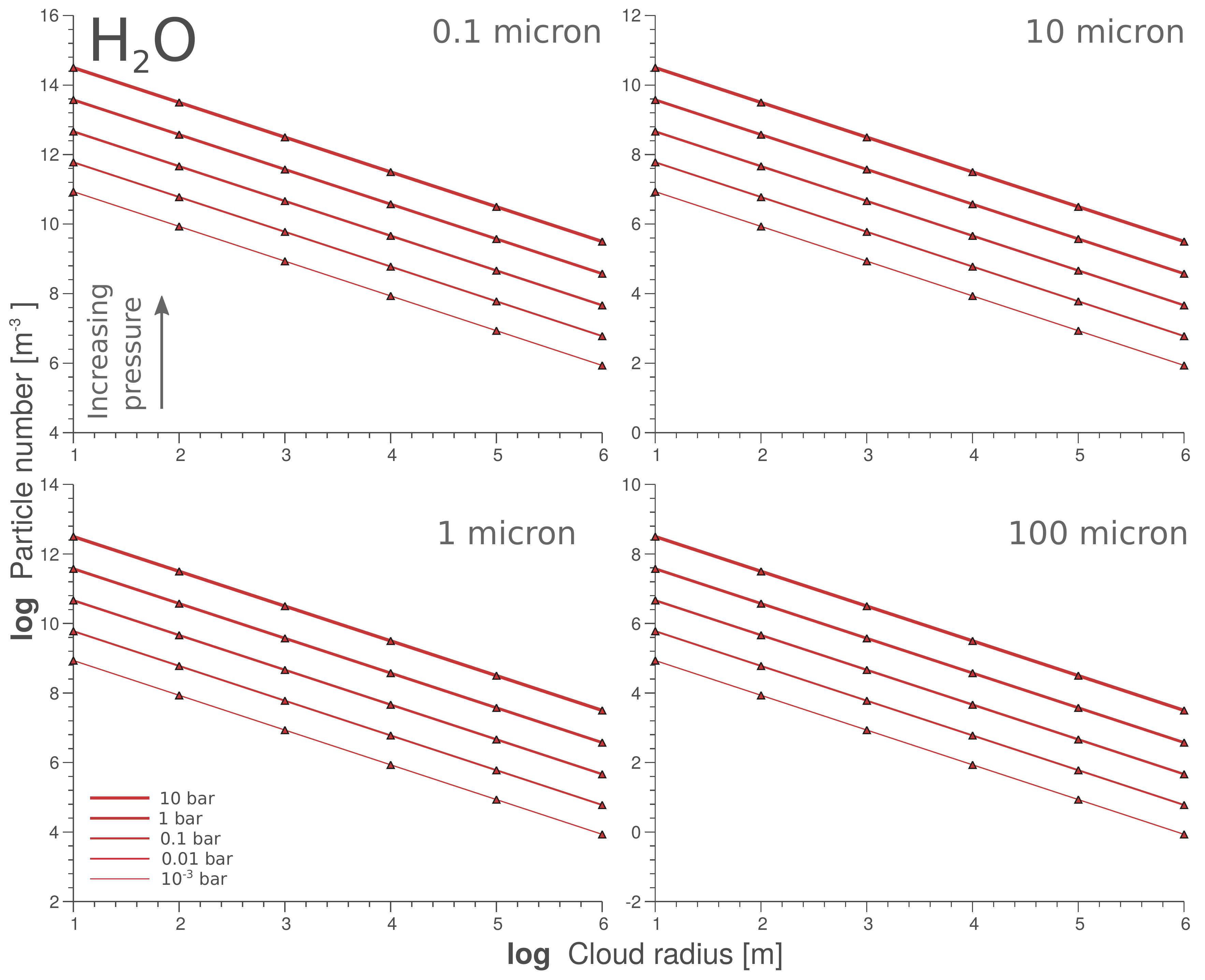}
        \caption{Same as figure \ref{fig_10_8} for a water atmosphere.}
        \label{fig_10_9}
\end{figure} 

\begin{figure}
    \centering
        \includegraphics[width=\linewidth, clip]{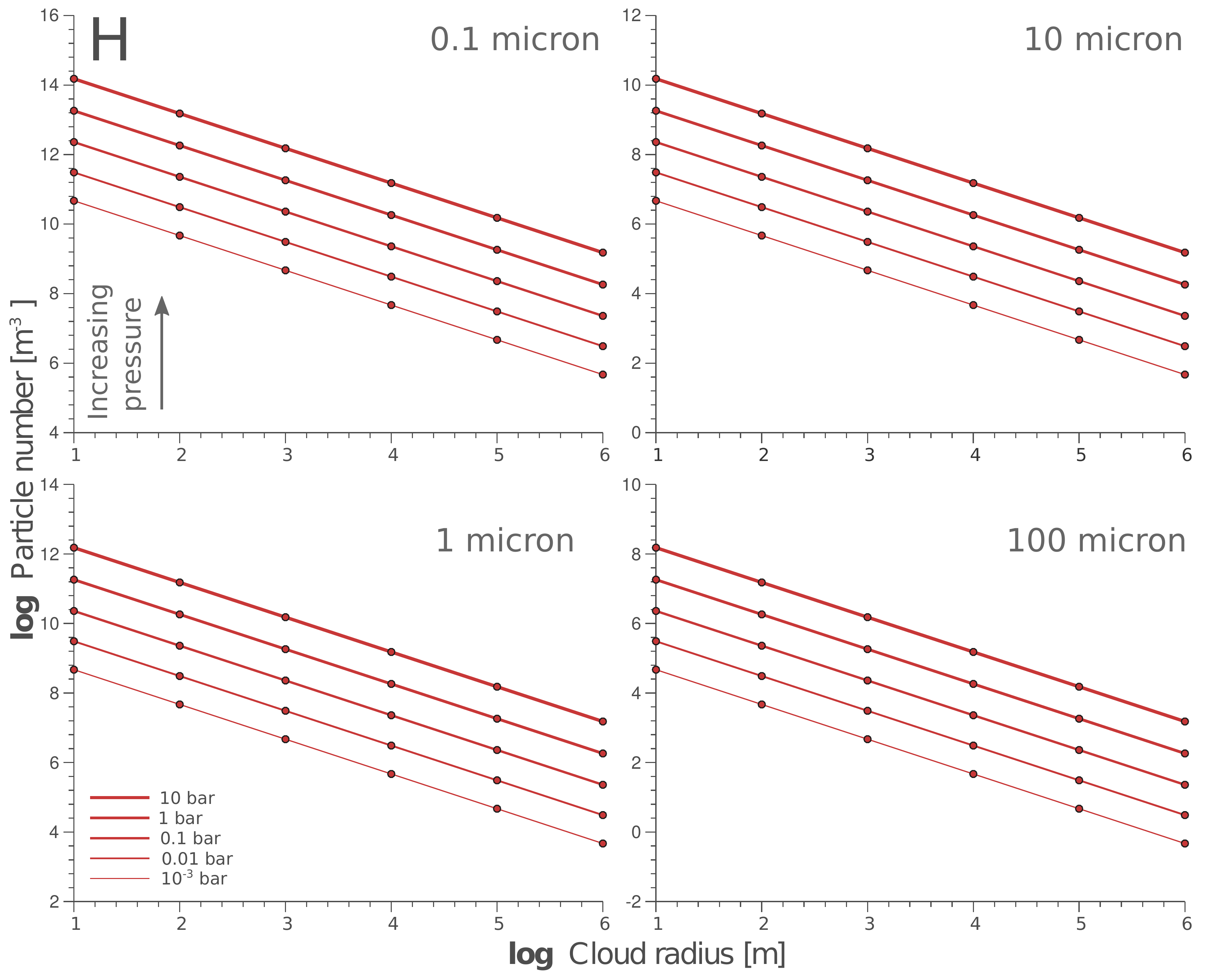}
        \caption{Same as figure \ref{fig_10_8} for a hydrogen atmosphere.}
        \label{fig_10_10}
\end{figure} 

\begin{figure}
    \centering
        \includegraphics[width=\linewidth, clip]{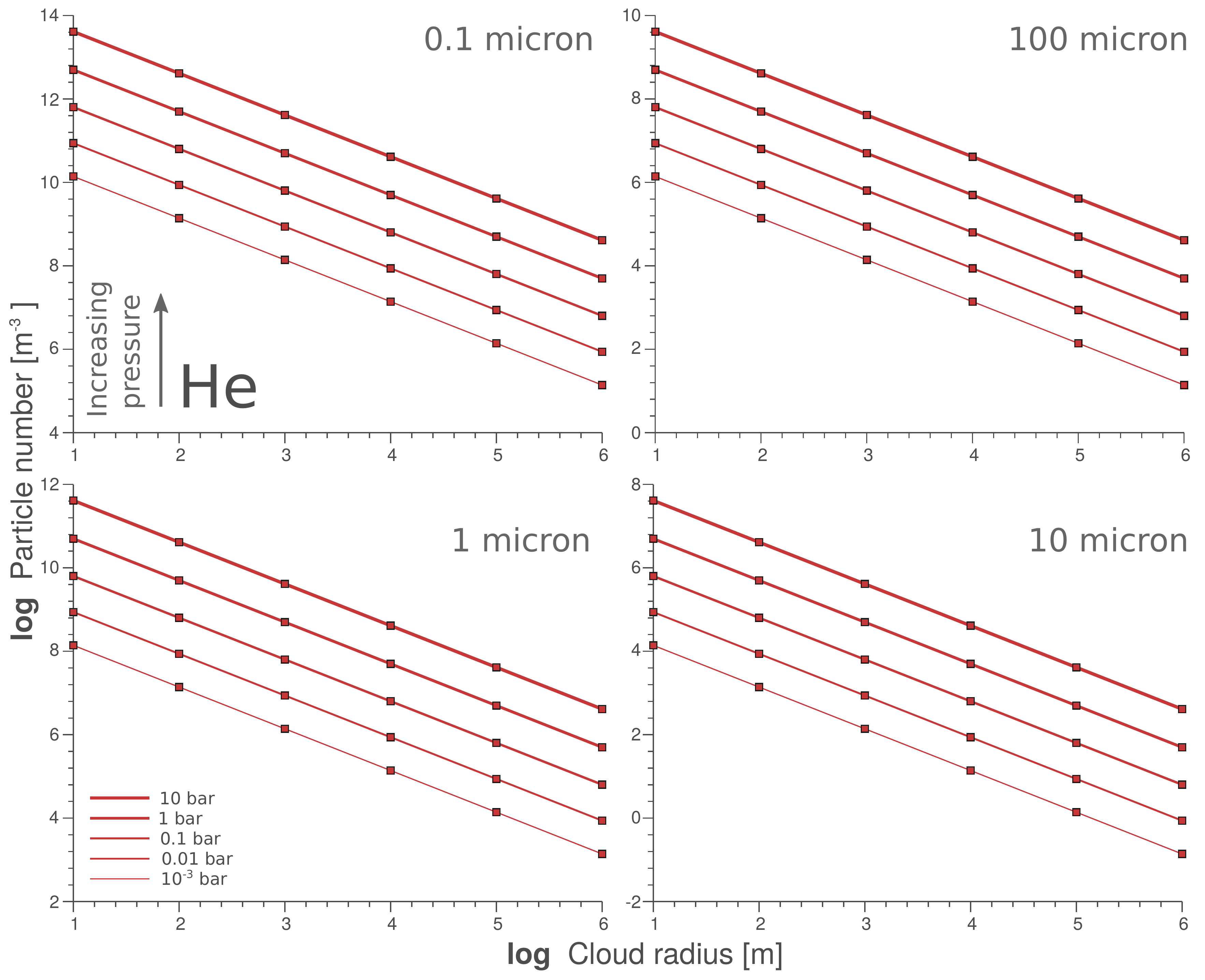}
        \caption{Same as figure \ref{fig_10_8} for a helium atmosphere.}
        \label{fig_10_11}
\end{figure} 

Based on the charge densities obtained from our experiments and the insight gained from equation \ref{eq10.6}, we proceed to determine the cloud conditions (cloud radius, particle density, and  size) required to produce the large gap breakdown field (at the cloud surface) for the four different gases we have been considering. The combination of cloud parameters that lead to discharge conditions, assuming that grains carry the maximum charge densities observed in the experiments (4 $\times$ 10\textsuperscript{-6} C m\textsuperscript{-2}), are shown in figures \ref{fig_10_8}b, \ref{fig_10_9}, \ref{fig_10_10}, and \ref{fig_10_11} for air, H\textsubscript{2}O, H, and He, respectively. Each of the four figures encompasses four panels corresponding to four different grain sizes (0.1, 1, 10, 100 microns). In turn, each panel renders the number of particles required to generate $E_b(p)$ as a function of cloud radius and pressure (four different curves). As expected, for a given cloud radius, clouds with larger particles require fewer particles to reach breakdown conditions at the surface of the cloud. Conversely, clouds with very small particle sizes, necessitate comparatively high particle concentrations to generate electric fields at the cloud surface that exceed $E_b$. Figures \ref{fig_10_8}b through \ref{fig_10_11} also show that pressure (through Paschen's Law) strongly influences the conditions under which electrostatic discharges may be generated by KCl or ZnS clouds. For a cloud with a given radius, clouds at higher pressures must harbor many more particles than clouds in more rarefied environments. 
In the particular case of GJ 1214b, \cite{charnay20153d} and \cite{gao2017clouds} indicate that the super-Earth maintains clouds at pressures in the vicinity of 0.1 bar composed of particles with diameters of 0.1 to 10 microns. Furthermore, modeling by \citep{gao2017clouds} suggest that these clouds  would have maximum number densities in the range of 10\textsuperscript{4} - 10\textsuperscript{5} m\textsuperscript{-3}. Assuming grains carry charge densities on the order of 10\textsuperscript{-6} C m\textsuperscript{-2} (the maximum charge densities observed in our experiments) clouds on GJ 1214b would have to have radii between 10 and 100 km to produce discharges. 

It is important to note that the analysis above only considers discharge processes resulting from the direct dielectric failure of the gas. When charge is distributed across many surfaces in large volumes, alternate processes may allow for discharges under weaker electrical stresses. Dispersed particles in a gas, for instance, tend to lower the breakdown field by acting as lenses that concentrate electric field lines at their surfaces \citep{cookson1970particle, schroeder1999model,solomon2001lightning}. Such focusing becomes more important as the contrast between the atmospheric dielectric constant and that of the grains increases. In thunderstorms, for instance, the maximum electric field may be reduced by a factor of 3 in the presence of particles. Furthermore, in large systems, discharge may be controlled by mechanisms that are not present at smaller scales. On Earth, for example, the electric fields in lightning-producing clouds rarely, if ever, exceed values of ~10\textsuperscript{5} Vm\textsuperscript{-1} (corrected for altitude) which is one order of magnitude smaller than the conventional breakdown field of 3 $\times$ 10\textsuperscript{6} Vm\textsuperscript{-1} predicted by Paschen's Law  \citep{marshall_electric_1995, schroeder1999model, solomon2001lightning, dwyer_initiation_2005}. In these regimes, the initiation of discharges in clouds may be controlled by a runaway electron avalanche, rather than the dielectric breakdown of the gas \citep{gurevich_runaway_1992, dwyer_initiation_2005, dwyer_physics_2014}. Runaway electrons are produced when the electric field in a charged particle cloud is large enough give electrons generated by cosmic rays the energy needed to overcome the losses associated with collisions and accelerate them to relativistic speeds. These electrons then collide with neutrals to generate more electrons that eventually lead to breakdown. In terrestrial environments, these processes likely occur not only in conventional thunderclouds but in other large-scale multiphase systems like as volcanic plumes and dust storms \citep{mcnutt_volcanic_2010}. If similar runaway avalanches are active in the putative clouds of exoplanets, grains would not be required to carry charge densities as high as the Paschen limit to produce large spark discharges. The environmental constraints presented in figures \ref{fig_10_8}b through \ref{fig_10_11} could be relaxed. However, outside of this discussion, such considerations are beyond the scope of this work. 
         
\section{Conclusions} 

Given the electrical diversity observed in our own solar system, an exoplanet with a dynamic, particle-bearing atmosphere (such as that inferred to exist on GJ 1214b) will almost certainly be characterized by complex electrostatic processes \citep{helling_ionization_2013, vorgul2016flash}. In this work, we have characterized the triboelectric behavior of potassium chloride and zinc sulfide under temperatures relevant to a number of exoplanets. The charge densities observed in our experiments are consistent with those measured on lightning-producing volcanic ash particles. Furthermore, we discussed the conditions under which such charge densities could lead to breakdown in mineral clouds. Together, these results provide further evidence that electrified salt grains in the atmospheres of distant worlds could drive discharge processes, potentially catalyzing prebiotic chemistry. Whether or not such activity could be detected on an exosolar object, nonetheless, remains contentious subject. In principle, lightning on worlds with charged, dusty atmospheres may be detected through excess radio emissions observed during transit \citep{vorgul2016flash}, or through abnormal spectral signals modulated by lightning-induced disequilibrium chemistry \citep{bailey2014ionization, xue2015spectral}. \cite{hodosan2016lightning} has suggested that a world's daytime spectrum could be cross-correlated with known lightning spectra to infer the presence of large-scale discharges. Yet, even for a world with an extremely high flash-rate, such signals may be too tenuous to be detected with even a state-of-the-art instrument like James Webb Space Telescope\citep{ardaseva2017lightning}. Definite detection of electrostatic processes may have to wait until the next generation of telescopes such as the European Extremely Large Telescope come on-line.
\acknowledgments
Acknowledgments. We would like to thank Dr. Laura Kreidberg for her helpful conversations on the clouds of GJ 1214b and Julian McAdams for his help preparing materials. This research was supported by the Blue Waters Graduate Fellowship.
\bibliography{references.bbl}
\end{document}